\documentclass[preprint,aps,12pt,preprintnumbers,eqsecnum,nofootinbib]{revtex4}
\usepackage{graphicx}
\usepackage{subfigure}

\usepackage{color}
\usepackage{amssymb,amsmath}

\begin{document}
%
%
\title{\vspace*{0.5in} Pseudo-Familon Dark Matter
\vskip 0.1in}
\author{Christopher D. Carone}\email[]{cdcaro@wm.edu}
\affiliation{High Energy Theory Group, Department of Physics,
College of William and Mary, Williamsburg, VA 23187-8795}
\date{December 2011}
\begin{abstract}
Motivated by a model of pseudo-Majoron dark matter, we show how the breaking of a  global symmetry that  acts nontrivially in lepton 
generation space can lead to a viable pseudo-familon dark matter candidate.  Unlike the pseudo-Majoron, the pseudo-familon in our model decays 
primarily to charged leptons and can account for the excess observed in the cosmic ray electron and positron spectra.
\end{abstract}
\maketitle

\section{Introduction} \label{sec:intro}

The compelling evidence for non-baryonic dark matter has motivated a vast literature on possible dark matter candidates.   Successful models must lead to the correct relic density, as required, for 
example, by the observed rotation curves of galaxies.  However, depending on what other phenomena one wishes to explain by the presence of dark matter, the details of viable scenarios can differ substantially 
from model to model.  In the present work, the phenomena we will attribute to the presence of dark matter is the striking rise in the cosmic ray positron fraction, from approximately $7$ to $100$~GeV observed in the  
PAMELA experiment~\cite{Adriani:2008zr} and the apparent excess in the total electron and positron flux, up to about $\sim 1$~TeV, as measured by Fermi-LAT~\cite{Abdo:2009zk} and H.E.S.S.~\cite{Aharonian:2009ah}.   
While the discrepancy between the observed spectra and predicted backgrounds might have a purely astrophysical origin (for example, if nearby pulsars are present~\cite{Hooper:2008kg, Yuksel:2008rf}),  dark matter 
annihilation~\cite{anex} or decays~\cite{deex} can also provide a source for the excess electron and positron flux.  (A recent review of the literature can be found in Ref.~\cite{Fan:2010yq}.)  In models with dark matter that is 
not exactly stable, fits to the data suggest that the dark matter must decay primarily to leptons and have a lifetime of ${\mathcal O}(10^{26})$~s.

Among the many interesting proposals for dark matter candidates in the recent literature is one in which a pseudo-Majoron is identified as the dark matter candidate~\cite{gms}. The Majoron is a goldstone boson of 
spontaneously broken lepton number and the pseudo-Majoron is the pseudo-goldstone boson that is obtained when the same spontaneously broken global symmetry is also explicitly broken by a small amount. 
In the model of Ref.~\cite{gms}, the explicit breaking is accomplished via soft terms in the scalar potential, and the sense in which this breaking is small is that the mass scale of these terms is much smaller than the scale
at which lepton number is spontaneously broken.  Although the Majoron can decay to standard model neutrinos, the lifetime is extremely long, due to the high-scale associated with the breaking
of the global symmetry.  While we will argue that the soft symmetry-breaking terms assumed in the original pseudo-Majoron proposal do not lead to a viable model, we also show how this difficulty can be overcome.
Hence, the Majoron can serve as a dark matter candidate and perhaps also provide detectable signals in neutrino telescopes.

The model of Ref.~\cite{gms}, however, is not one which can account for the cosmic ray electron and positron excess.  Here, we will show that it is possible to construct a similar model in which 
the dark matter candidate is also the pseudo-goldstone boson of a spontaneously broken approximate global symmetry, but one which leads to couplings of the dark matter candidate to charged leptons rather than 
neutrinos.  Unlike lepton number (which treats all generations of leptons identically) our global symmetry will not.  Hence, the dark matter candidate that we propose here is technically a pseudo-familon, at least 
as far as the lepton sector is concerned. The model we discuss in the next sections provides an existence proof that pseudo-familon dark matter is a viable possibility that can accommodate 
the cosmic ray data.

\section{The Model}
We assume the existence of global symmetry $G=\mbox{U(1)}$ that acts nontrivially in lepton generation space.  The charge assignments of the fields are summarized in Table~\ref{table:one}. 
\begin{table}
\begin{tabular}{lcccccccccccccc} \hline\hline
Fields:       & \hspace{0.2em}&$L_L^{1,2}$ & \hspace{0.2em} & $L_L^{3}$ & \hspace{0.2em} & ${e_R}^{1,2}$ &  \hspace{0.2em}&${e_R}^{3}$ & \hspace{0.2em}& $\nu_R^{i}$ & \hspace{0.2em}& $\phi$ & \hspace{0.2em} &$H$ \\ \hline
Charges:  & &$+4$ & & $-4$ & & $+4$ & & $-4$ & & $0$ & & $1$ & &$0$ \\ \hline\hline
\end{tabular}
\caption{Charge assignments under the U(1) global symmetry.  The superscripts indicated generation number.}\label{table:one}
\end{table}
Notice that the global symmetry distinguishes charged leptons of the third generation from those of the other two.  The complex scalar field $\phi$
has charge $+1$ and its vacuum expectation value (vev) completely breaks the symmetry
\begin{equation}
\mbox{U(1)}\stackrel{\langle \phi \rangle}{\longrightarrow}\,\mbox{nothing}  \, .
\label{eq:brk}
\end{equation} 
The standard model Higgs doublet $H$ is neutral under $G$ and plays its conventional role in electroweak symmetry breaking and the generation of fermion masses.

The spontaneous breaking of the global symmetry leads to a goldstone boson $\varphi$, which we shall identify via the conventional nonlinear field redefinition
\begin{equation}
\phi = \frac{v_\phi + \sigma}{\sqrt{2}} \exp[ i \varphi/v_\phi] \,\,\,  ,
\label{eq:nonlin}
\end{equation}
where $\langle \phi \rangle \equiv v_\phi/\sqrt{2}$. We will show that when one includes an appropriate set of terms that softly break $G$, the familon $\varphi$ becomes massive and can annihilate via couplings that are not 
suppressed by the scale $v_\phi$; regions of the model parameter space exist where the familon is sufficiently long lived and has the necessary relic density to be a viable dark matter candidate.

Before proceeding to evaluate the model that we have just defined, it is worth clarifying some of the underlying assumptions.  First, we assume the existence of a global symmetry.  While such symmetries can be 
violated by generic quantum gravitational effects, it is also true that they can arise accidentally in low-energy effective theories as a consequence of the gauge symmetries in an ultraviolet completion.   Hence, as in the Majoron model of Ref.~\cite{gms}, it is  reasonable to consider the consequences of relatively simple global symmetries below the Planck scale that lead to interesting phenomenology.  Secondly, we employ the term familon since  our global symmetry 
$G$ acts nontrivially in lepton generation space.  This does not imply, however, that $G$ is intended to provide a complete understanding of fermion Yukawa textures by itself.  Instead, we require that $G$ not restrict the lepton Yukawa textures too severely, so that the present construction is compatible with a wide range of possible flavor models; these may introduce additional symmetries in generation space.  Finally,  the model we present is non-supersymmetric,  as a matter of convenience.  Possible solutions to the problem of fine-tuning in theories with fundamental scalars are well known, but are largely independent of the issues that we consider here.    Depending on what we  learn from the Large Hadron Collider, one can modify the model accordingly.  

The symmetry $G$ partly restricts the lepton Yukawa couplings.  The charged lepton Yukawa matrix $Y_L$, defined by
\begin{equation}
-{\cal L} \supset  \bar{L}_L Y_L H e_R + \mbox{h.c. } \, ,
\end{equation}
indicates that $Y_L$ must have the global charges
\begin{equation}
Y_L \sim \left( \begin{array}{cc|c}
0 & 0 & 8 \\
0 & 0 & 8 \\ \hline
-8 & -8 & 0
\end{array} \right) \, .
\end{equation}
The entries with non-vanishing charges arise via higher-dimension operators, so that at lowest order
\begin{equation}
Y_L = 
 \left( \begin{array}{cc|c}
a_{11} & a_{12} & 0  \\
a_{21} & a_{22} & 0  \\ \hline
0 & 0 & a_{33}  \end{array} \right)+
\frac{1}{\Lambda^8}  \left( \begin{array}{cc|c}
0 & 0 & b_{13} \, \phi^{8}  \\
0 & 0 & b_{23} \, \phi^{8} \\ \hline
b_{31} \, \phi^{*8} & b_{32} \, \phi^{*8} & 0 \end{array} \right)  \, ,
\end{equation}
where the $a_{ij}$ and $b_{ij}$ are coefficients that are at most ${\cal O}(1)$, and $\Lambda$ is the lepton flavor scale, the scale at which the higher-dimension operators are generated.  This form clearly can parametrically 
accommodate the charged lepton Yukawa couplings.  However,  one can do better: the breaking of additional symmetries can provide an explanation for the charged lepton mass hierarchy.  For 
example, consider a $Z_{6}$ flavor symmetry under which the three generations of right-handed charged leptons transform as  $e_R^{(i)} \rightarrow \omega^{i+2} e_R^{(i)}$, for $i=1\ldots3$, 
where  $\omega^{6}=1$.  Then given a flavon field $\phi_F \rightarrow \omega \phi_F$, one would find instead
\begin{equation}
Y_L = 
\left( \begin{array}{cc|c}
a_{11} \phi_F^3/\Lambda^3 & a_{12} \phi_F^2/\Lambda^2 & 0  \\
a_{21} \phi_F^3/\Lambda^3 & a_{22} \phi_F^2/\Lambda^2 & 0  \\ \hline
0 & 0 & a_{33}  \phi_F/\Lambda \end{array} \right)+
\left( \begin{array}{cc|c}
0 & 0 & b_{13} \, \phi_F \phi^{8}/\Lambda^9 \\
0 & 0 & b_{23} \, \phi_F \phi^{8}/\Lambda^9 \\ \hline
 b_{31} \, \phi_F^3 \phi^{*8}/\Lambda^{11} & b_{32} \,\phi_F^2  \phi^{*8}/\Lambda^{10} & 0 \end{array} \right)  \, .
\end{equation}
Assuming $\langle \phi_F \rangle \approx \langle \phi \rangle \approx \lambda^2\, \Lambda$, where $\lambda \approx 0.2$ is of order the Cabibbo angle, one obtains the viable charged lepton
Yukawa texture
\begin{equation}
Y_L = \left( \begin{array}{cc|c}
 a_{11} \,  \lambda^6 & a_{12} \, \lambda^4 &  b_{13}\, \lambda^{18} \\
a_{21} \, \lambda^6 & a_{22} \, \lambda^4 & b_{23}\, \lambda^{18}  \\ \hline 
b_{31} \, \lambda^{22} & b_{32}\, \lambda^{20} & a_{33}\, \lambda^2 \end{array} \right)  \, ,
\label{eq:texture}
\end{equation}
and the $\varphi \, \ell^+ \ell^-$ coupling
\begin{equation}
{\cal L} \supset 4 i \sqrt{2} \frac{\langle H \rangle}{\langle \phi \rangle} \varphi \,\bar \ell_L
 \left( \begin{array}{cc|c}
 0 & 0 & b_{13}\, \lambda^{18} \\
0 & 0 & b_{23}\, \lambda^{18}  \\ \hline 
b_{31} \, \lambda^{22} & b_{32} \,  \lambda^{20} & 0 \end{array} \right) \ell_R  + \mbox{ h.c. }\,,
\label{eq:cmatrix}
\end{equation}
where $\ell = (e,\mu,\tau)^T$. Note that the vanishing entries in Eq.~(\ref{eq:cmatrix}) are not corrected at higher order.  Since the diagonal blocks of Eq.~(\ref{eq:texture}) are 
neutral under $G$, any higher-order corrections to these entries that involve $\phi$ must do so via the product $\phi^\dagger \phi$, which is independent of $\varphi$.  Note also that the matrices 
in Eqs.~(\ref{eq:texture}) and (\ref{eq:cmatrix}) are not proportional to each other and cannot be diagonalized simultaneously.   Due to the extreme smallness of the $13$ and $23$ rotation angles required 
to diagonalize $Y_L$, one finds that the largest entries of the familon coupling matrix are the same as in Eq.~(\ref{eq:cmatrix}), though generally with different order-one coefficients.  In the mass eigenstate 
basis, one finds
\begin{equation}
{\cal L} \supset 4 i \sqrt{2} \frac{\langle H \rangle}{\langle \phi \rangle} \varphi \,\bar \ell_{0L}
 \left( \begin{array}{cc|c}
 {\cal O}(\lambda^{38}) & {\cal O}(\lambda^{36}) & B_{13}\, \lambda^{18} \\
{\cal O}(\lambda^{38}) & {\cal O}(\lambda^{36}) & B_{23}\, \lambda^{18}  \\ \hline 
{\cal O}(\lambda^{22}) & {\cal O}(\lambda^{20}) & {\cal O}(\lambda^{34}) \end{array} \right) \ell_{0R}  + \mbox{ h.c. }\, ,
\label{eq:cmbasis}
\end{equation}
where we have only retained order-one coefficients $B_{ij}$ for the non-negligible terms.  The $\varphi$ decays are lepton flavor violating; the leading decay channels are
$\varphi \rightarrow e^- \tau^+$, $e^+ \tau^-$, $\mu^- \tau^+$ and $\mu^+ \tau^-$.  The dark matter lifetime of ${\cal O}(10^{26})$~s  is 
obtained when the operator coefficient satisfies
\begin{equation}
4\sqrt{2} \frac{\langle H \rangle}{\langle \phi\rangle} \, \lambda^{18} \sim 4 \sqrt{2} \frac{\langle H \rangle}{\Lambda} \lambda^{16} \sim 10^{-26} \, ,
\label{eq:Flifetime}
\end{equation}
which implies  $\Lambda \sim 10^{18}$~GeV, up to order one uncertainty.  We will take $\Lambda \sim M_*$, the reduced Planck mass, henceforth.  One should keep in mind, however,
that the scale $\Lambda$ could arise via renormalizable physics, for example, the integrating-out of vector-like fermions with masses just below $M_*$.

In the pseudo-Majoron dark matter model, the Majoron decay amplitude is proportional to neutrino masses, which accounts for the longevity of the dark matter candidate. One might 
naively expect that a generic familon in the charged lepton sector would decay too quickly, due to the much larger charged lepton masses, to be a viable dark matter candidate.  The crucial 
feature of the model we have presented is that our familon couplings are proportional to {\em off-diagonal} elements of the charged lepton mass matrix. These elements can be arbitrarily small,
leading to the long dark matter lifetime required by the cosmic ray data. 

Neutrino masses in the model arise via the see-saw mechanism.   The Dirac and Majorana mass terms are defined by
\begin{equation} 
- {\cal L} \supset  \bar \nu_R^c  M_{RR}  \, \nu_R +  (\bar L_L Y_{LR} H \nu_R + \mbox{ h.c. } )  
\end{equation}
The entries of $Y_{LR}$ have the global charges 
\begin{equation}
Y_{LR}  \sim \left( \begin{array}{cc|c} 4 & 4 & 4 \\
4  & 4 & 4 \\ \hline
-4 & -4 & -4 \end{array} \right)
\end{equation}
and hence arise all at the same order $\langle \phi \rangle^4/\Lambda^4 \sim \lambda^8$; we may therefore write  $Y_{LR} = \lambda^8 \tilde{Y}_{LR}$, where $\tilde{Y}_{LR}$ is
an arbitrary matrix with order-one entries.  The matrix $M_{RR}$ is not restricted at all by $G$.  Hence, the flavor structure of the neutrino sector is largely unconstrained 
and can be chosen to accommodate the data. On the other hand, the overall light neutrino mass scale, $m_\nu$, follows from the seesaw formula 
\begin{equation}
m_\nu \sim \frac{\lambda^{16} \langle H \rangle^2}{M_R}
\end{equation}
where $M_R$ characterizes the right-handed neutrino mass scale.  Choosing $m_\nu \sim 0.05\mbox{ eV}$  (which is compatible with the data on atmospheric neutrino oscillations~\cite{pdg}), one
finds $M_R \sim {\cal O}(10)$~TeV.    This scale is high enough so that the see-saw formula is an accurate approximation and the heavy, right-handed neutrino mass eigenstates present  no 
phenomenological problems.

\section{Scalar Couplings}

Let us first consider the U(1) invariant portion of the scalar sector of the theory.   Imagine we employ the linear decomposition $\phi = (v_\phi + \phi_r + i\, \phi_i)/\sqrt{2}$, where $\phi_i$ is the massless
degree of freedom.   Let $H$ represent the standard model Higgs doublet, and $h$ the Higgs boson.   In the pseudo-Majoron model of Ref.~\cite{gms}, it was argued that the coupling 
$\phi^\dagger \phi H^\dagger H$ leads to interactions of the form $h^2 \phi_i^2$ and $v\, h\,  \phi_i^2$, where $v/\sqrt{2} \equiv \langle H \rangle$.   Since these are the standard Higgs portal couplings 
for a scalar dark matter candidate, the authors of Ref.~\cite{gms} argued that the appropriate relic density could readily be achieved.

This argument, however, is not correct.   Using the nonlinear redefinition of $\phi$, Eq.~(\ref{eq:nonlin}), one notices that the U(1) invariant $\phi^\dagger \phi$ is independent of the
goldstone mode $\varphi$;  hence, one concludes using this representation that there are {\em no} couplings of $\varphi$ to $h$ that are unsuppressed by the scale $v_\phi$.  Both the linear and non-linear representations of the 
goldstone boson, however,  should lead to the same physical results.   This puzzle can be resolved in the linear theory by carefully evaluating
the scalar couplings in the mass eigenstate basis.  For example, one finds couplings of the form  $v \, h\,  \phi_i^2$ and $v_\phi \, \phi_r \, \phi_i^2$, where $\phi_r$ is (mostly) a super-heavy scalar state with mass 
of order $v_\phi$.   One cannot simply neglect the coupling to $\phi_r$: there is mixing between $h$ and $\phi_r$ such that $h \sim h_0$ and $\phi_r \sim {\cal O}(v/v_\phi) \, h_0$, where $h_0$ is the Higgs mass 
eigenstate.   Thus, one finds  {\em two} contributions to the $h_0 \, \phi_i^2$ vertex, each with a coefficient of ${\cal O}(v)$.   It is straightforward to verify that they cancel, up to corrections of  ${\cal O}(v^3/v_\phi^2)$.
The absence of couplings that are unsuppressed by the high scale $v_\phi$ implies that one cannot achieve a sufficient annihilation cross section via the couplings in the U(1) invariant sector of the theory.

The previous observations suggest that the desired Higgs-portal couplings must arise in the soft-symmetry breaking sector of the theory.   As in Ref.~\cite{gms}, we assume that the global symmetry is an approximate one 
due to the existence of renormalizable interactions with mass dimension three or less that explicitly break the symmetry.   Such interactions must provide for an adequate dark matter mass and annihilation cross section, 
while not leading to rapid dark matter decays.    To obtain the dark matter mass scale that is suggested by the cosmic-ray data, we take the new dimensionful couplings to be TeV-scale in size. 
The familon $\varphi$ is nonetheless correctly categorized as a pseudo-goldstone boson since the ratio of the scales of explicit to spontaneous symmetry breaking is  small,  $(10^3 \mbox{ GeV})/(10^{17}\mbox{ GeV} )\sim 10^{-14}$. 
The simplest way to allow for unsuppressed couplings that lead to dark matter annihilation is to introduce a real scalar singlet $S_0$ and the CP-invariant symmetry-breaking terms $m_1^2 \phi^2 + m_2 \, S_0 \, \phi^2 + \mbox{ h.c.}$   One may 
consistently impose a $Z_2$ symmetry that eliminates soft terms that are odd in $\phi$, while allowing all the other Lagrangian terms that we have considered thus far.   Note that $m_1$ can always be made real by a global U(1) rotation, while 
$m_2$ is taken real as a parameter choice; this satisfies `t Hooft's criterion for technical naturalness since the scalar sector then has an enhanced symmetry, namely a $Z_2$ where $\varphi \rightarrow -\varphi$.  It follows that the scalar-sector 
interactions involve only even powers of pseudo-goldstone boson field and therefore do not introduce an avenue for rapid decays~\footnote{Of course, one may obtain scalar decay channels via diagrams involving the extremely
small $\varphi \, \ell^+ \ell^-$ couplings which violate this $Z_2$.  However, these decays can be neglected since their branching fractions are loop suppressed relative to the leading modes discussed earlier.}.  The term proportional to $m_2$ 
includes the interaction $m_2 S \, \varphi^2/2$, where $S$ represents the fluctuation of the singlet about its vev.    Provided that the necessary tunings of dimensionful couplings are employed so that the scalar $S$ and the Higgs field $h$ are 
light, then the low-energy effective theory will contain the terms
\begin{equation}
V \supset \frac{1}{2} m_\varphi^2 \, \varphi^2 + \frac{1}{2} m_S^2 S^2 + \frac{1}{2} m_2 S \varphi^2 + \frac{1}{2} \mu S h^2 + \cdots  \,\,\, ,
\label{eq:effv}
\end{equation}
where the dimensionful parameters shown are TeV-scale in size.  Generically, $S$ and $h$ will mix, so that the familon can annihilate to standard model particles via the $s$-channel exchanges of
both mass eigenstates.  A complete study of the allowed parameters space that yields the correct relic density is beyond the scope of his letter (a detailed study of a similar sector can be found in Ref.~\cite{cep}).
Here we will focus on a well motivated limit:  For a Higgs boson that is very similar to that of the standard model, the $h$-$S$ mixing should be small.  If the singlet is an order of magnitude heavier than the Higgs 
({\em  i.e.}, similar in mass to the familon), then we would expect this to be a reasonable approximation.    With mixing effects neglected,  one sees that the interactions in Eq.~(\ref{eq:effv}) nonetheless 
lead to the dark matter annihilation channel $\varphi \varphi \rightarrow h h$, via $s$-channel exchanges of the singlet $S$.  The annihilation cross section is given by
\begin{equation}
\sigma(\varphi\varphi \rightarrow h h) = \frac{1}{32 \pi} \frac{1}{s} \left[\frac{s - 4 m_h^2}{s-4 m_\phi^2}\right]^{1/2} \frac{m_2^2 \, \mu^2}{(s-m_S^2)^2+m_S^2 \Gamma_S^2} \,\,\, ,
\end{equation}
where the singlet width, for $m_S < 2\, m_\varphi$, is given by
\begin{equation}
\Gamma_S (S \rightarrow hh) = \frac{1}{32 \pi} \left(1-\frac{4 m_h^2}{m_S^2}\right)^{1/2} \frac{\mu^2}{m_S}  \,\, .
\end{equation}
For example, for the parameter choices $m_\varphi = 2$~TeV, $m_S=3.5$~TeV, $m_2=2.5$~TeV, $\mu=2.0$~TeV and $m_h=125$~GeV, we find that the freeze out condition
\begin{equation}
\frac{ n_\varphi^{EQ} \langle \sigma v \rangle}{H(T_f)} \approx 1\, ,
\end{equation}
gives $x \equiv m_\varphi/T_f \approx 28$, where $n_\varphi^{EQ}$ is the equilibrium number density, $ \langle \sigma v \rangle$ is the thermally averaged annihilation cross section and $T_f$
is the freeze out temperature; we find that this leads in turn to the present dark matter density $\Omega_D h^2 \approx 0.1$, as desired\footnote{This analysis is quite standard; a more detailed discussion can be found in Ref.~\cite{fndm}.}.   
This example demonstrates that there are regions of the full parameters space of the model where the appropriate dark matter relic density can easily be obtained.

\section{Cosmic Rays}
\begin{figure}[t]
\subfigure{\includegraphics[width = 0.4\textwidth]{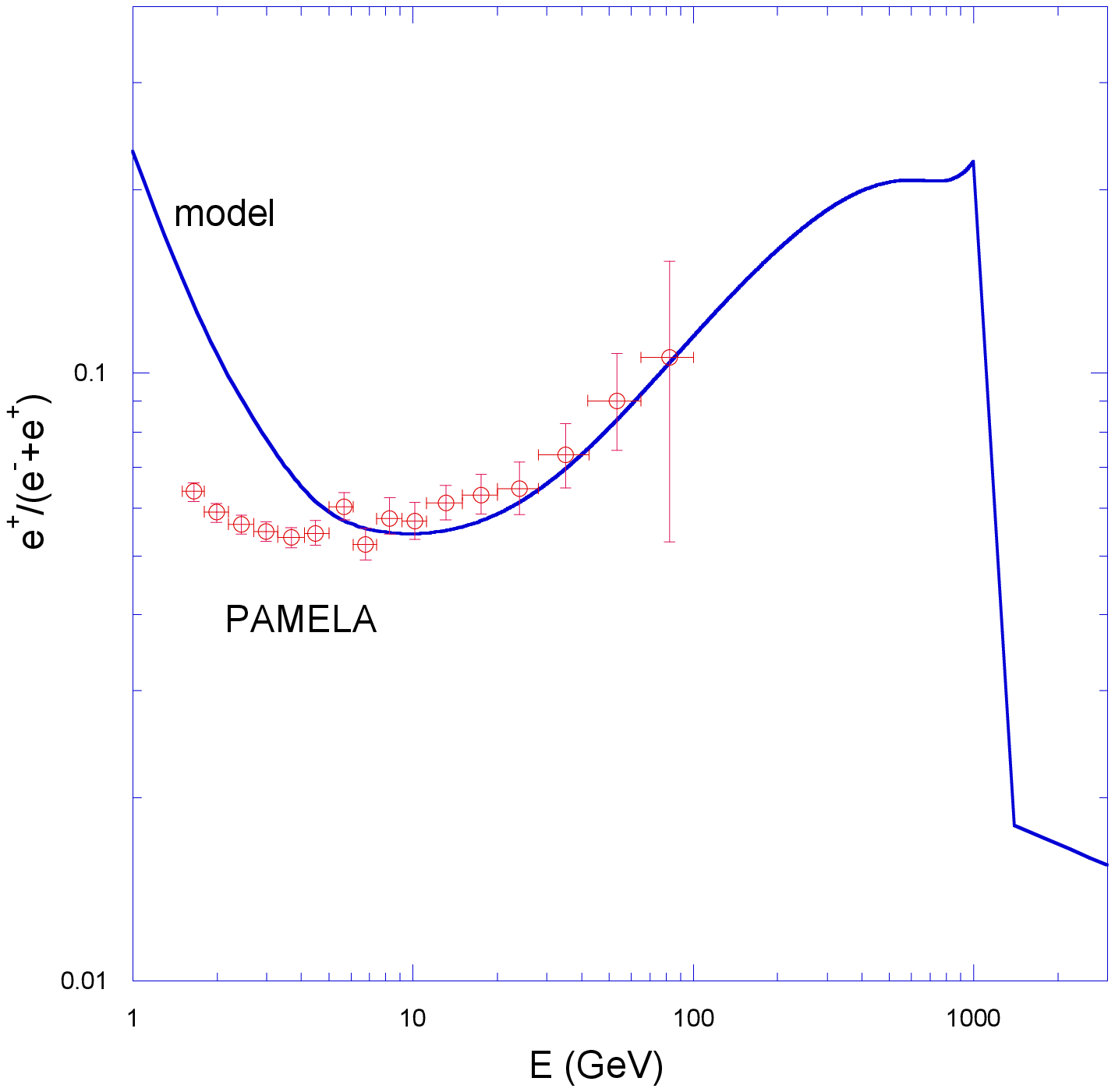}} \hspace{1em}
\subfigure{\includegraphics[width = 0.42\textwidth]{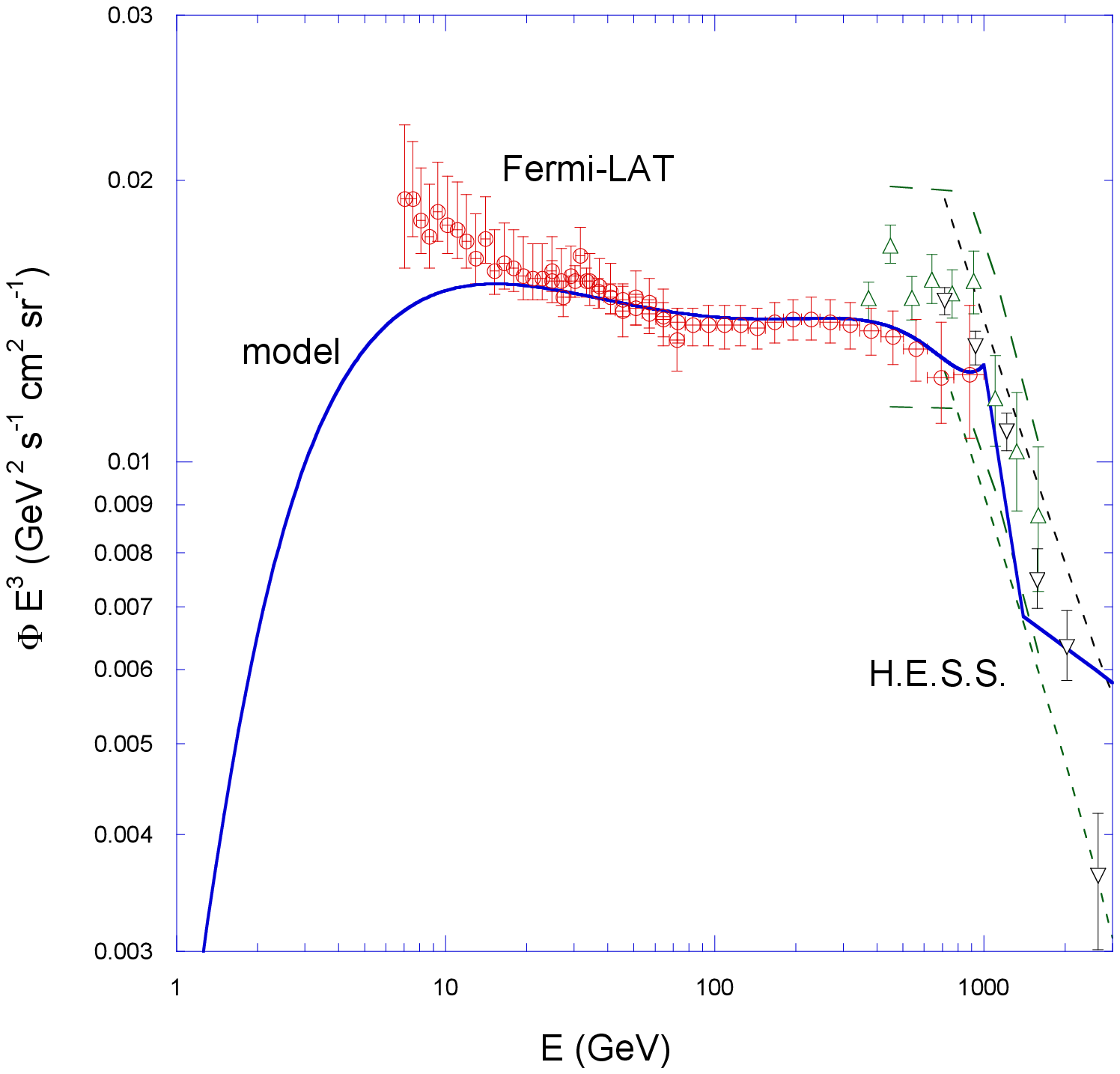}}
\caption{Positron fraction and total electron plus positron flux for the parameter choice $f=0.375$.  The best fit predicted spectrum ($\chi^2/\mbox{d.o.f}=0.61$) is 
shown, corresponding to the familon mass and lifetime $m_\varphi = 2$~TeV and $\tau_\varphi = 2 \times 10^{26}$~s, respectively.  Data from
PAMELA~\cite{Adriani:2008zr}, Fermi LAT~\cite{Abdo:2009zk} and H.E.S.S.~\cite{Aharonian:2009ah} are also shown.  The dashed lines indicate the H.E.S.S. band of systematic uncertainty.} \label{fig:one}
\end{figure}

The branching fractions for the leading familon decay channels are determined by the couplings $B_{13}$ and $B_{23}$ in Eq.~(\ref{eq:cmbasis}), i.e.,
\begin{equation}
{\cal L} \supset 4 i \sqrt{2} \, \frac{\langle H \rangle}{\langle \phi \rangle} \varphi \lambda^{18} \left[ B_{13} \bar{e} P_R \tau + B_{23} \bar\mu P_R \tau \right] + \mbox{ h.c. }  \,,
\end{equation}
where $P_R = (1+\gamma^5)/2$, and are given by
\begin{eqnarray}
B(e^- \tau^+) = B(e^+ \tau^-) &=& 0.5 \frac{|B_{13}|^2}{|B_{13}|^2 + |B_{23}|^2}  \nonumber \\
B(\mu^- \tau^+) = B(\mu^+ \tau^-) &=& 0.5 \frac{|B_{23}|^2}{|B_{13}|^2 + |B_{23}|^2} 
\label{eq:bfs}
\end{eqnarray}
Since the total must sum to one, we can parameterize Eq.~(\ref{eq:bfs}) in terms of the branching fraction to $\mu^- \tau^+$:  $B(\mu^- \tau^+) = B(\mu^+ \tau^-)=f$ and
$B(e^- \tau^+) = B(e^+ \tau^-)=1/2-f$, with $0 \leq f \leq 1/2$.   Hence the dark matter mass $m_\varphi$, lifetime $\tau_\varphi$ and the value of the parameter $f$
determines the contribution from dark matter decays to the cosmic ray spectra. We compute the energy distribution of cosmic ray electrons or positrons using PYTHIA 6.4~\cite{pythiaman}; when 
more than one decay channel is possible
\begin{equation}
\frac{dN_{e^{\pm}}}{dE} = \sum_{i,j} B( \ell^+_i \ell^-_j) \left( \frac{dN_{e^{\pm}}}{dE} \right)_{ij}  \, ,
\label{eq:totenspec}
\end{equation}
where $(dN_{e^{\pm}}/dE)_{ij}$ is the energy spectrum obtained given the primary decay  $\varphi \rightarrow \ell^+_i \ell^-_j$, with the subscript indicating lepton flavor.
The energy spectrum in Eq.~(\ref{eq:totenspec}) is then input into a standard diffusive model for propagation of electrons and positrons through the interstellar medium.  This
analysis is now standard to the literature on decaying dark matter and the details, including our assumptions on background fluxes, are identical to those found in Ref.~\cite{ari};
we refer the reader to this reference for a detailed discussion.  An example of typical results that can be obtained in the model are shown in Fig.~\ref{fig:one}.
Here we have fixed the parameter $f$ at $0.375$ and have performed a least-squares fit to determine the optimal values of $m_\varphi$ and $\tau_\varphi$.  Note that the ``step" in the
predicted total flux around $1$~TeV is due to the admixture of $e^\pm \tau^\mp$ in the primary decay, which contributes electrons and positrons of a fixed injection energy,
 $\approx m_\varphi/2$, to the spectrum. In the example shown, this feature is within the total experimental uncertainty of the data in this region, and can be made smaller
 for larger values of $f$.

\section{Conclusions}
Pseudo-Majoron dark matter is an interesting proposal that cannot account for the anomalies in the cosmic ray electron and positron spectra.  We have shown that the
breaking of a  global symmetry that acts nontrivially in lepton generation space can lead to a viable dark matter candidate that does decay primarily to charged
leptons.  The long lifetime of the  pseudo-familon dark matter candidate in the model we have proposed is related to the smallness of off-diagonal entries in the charged lepton Yukawa 
matrices;  no phenomenological consideration prevents these entries from being small, an outcome that is realized due to the symmetries, charge assignments and the choice
of mass scales in our model.  We have argued that unsuppressed Higgs-portal couplings can arise via soft symmetry-breaking terms involving a gauge singlet scalar, so that
the correct pseudo-familon relic density can be achieved.  We have shown also that the predicted cosmic ray $e^{\pm}$ spectra can fit  the current data from the Fermi-LAT, PAMELA 
and H.E.S.S experiments.  The model we have presented serves as a proof of principle; a more elegant implementation within a comprehensive framework for the origin of lepton flavor 
is a direction worthy of investigation.

\begin{acknowledgments}
This work was supported by the NSF under Grants PHY-0757481 and PHY-1068008.  C.D.C.\  thanks Josh Erlich for useful comments.  In addition, the author 
gratefully acknowledges support from a William \& Mary Plumeri Fellowship.
\end{acknowledgments}


\end{document}